\documentclass[aps,prl,superscriptaddress,footinbib,twocolumn]{revtex4-2}

\usepackage{dcolumn}
\usepackage{bm}

\usepackage[T1]{fontenc}
\usepackage{color}
\usepackage{graphicx}
\usepackage[american]{babel}

\usepackage{amsmath,amsthm,amsfonts,amssymb,amscd}
\usepackage{indentfirst}
\usepackage{float}
\usepackage[dvipsnames]{xcolor}
\usepackage[colorlinks=true,citecolor=ForestGreen,linkcolor=Mulberry,urlcolor=BlueViolet,hyperindex]{hyperref}
\usepackage{mathtools}
\usepackage{bbold}
\usepackage{units}
\usepackage{soul}
\usepackage{ulem}

\definecolor{gray}{rgb}{0.75,0.75,0.75}

\newcommand{\CASkey}{CAS Key Laboratory of Quantum Information, University of Science
and Technology of China, Hefei, 230026, China}
\newcommand{\CAScenter}{CAS Center For Excellence in Quantum Information and QuantumPhysics, Hefei, 230026, China}
\newcommand{\vcq}{University of Vienna, Faculty of Physics, Vienna Center for Quantum Science and Technology (VCQ) and Research platform TURIS, Boltzmanngasse 5, 1090 Vienna, Austria
}

\newcommand{\cdl}{Christian Doppler Laboratory for Photonic Quantum Computer, Faculty of Physics, University of Vienna, Sensengaasse 8, 1090 Vienna, Austria}
\newcommand{\nationallab}{Hefei National Laboratory, University of Science and Technology of China, Hefei 230088, China}

\begin{document}

\title{Experimental beating the standard quantum limit under non-markovian dephasing environment}

\author{Huan Cao}
\affiliation{\CASkey}
\affiliation{\vcq}
\affiliation{\CAScenter}
\affiliation{\cdl}

\author{Chao Zhang}
\email{drzhang.chao@ustc.edu.cn}
\affiliation{\CASkey}
\affiliation{\CAScenter}
\affiliation{\nationallab}

\author{Yun-Feng Huang}
\email{hyf@ustc.edu.cn}
\affiliation{\CASkey}
\affiliation{\CAScenter}
\affiliation{\nationallab}

\author{Bi-Heng Liu}
\affiliation{\CASkey}
\affiliation{\CAScenter}
\affiliation{\nationallab}

\author{Chuan-Feng Li}
\email{cfli@ustc.edu.cn}
\affiliation{\CASkey}
\affiliation{\CAScenter}
\affiliation{\nationallab}

\author{Philip Walther}
\affiliation{\vcq}
\affiliation{\cdl}

\author{Guang-Can Guo}
\affiliation{\CASkey}
\affiliation{\CAScenter}
\affiliation{\nationallab}


\begin{abstract}
Entanglement enhanced quantum metrology has been well investigated for beating the standard quantum limit (SQL). However, the metrological advantage of entangled states becomes much more elusive in the presence of noise. Under strict Markovian dephasing noise, the uncorrelated and maximally entangled states achieve exactly the same measurement precision. However, it was predicted that in a non-Markovian dephasing channel, the entangled probes can recover their metrological advantage. Here, by using a highly controlled photonic system, we simulate a non-Markovian dephasing channel that fulfills the quadratic decay behaviour. Under such a channel, we demonstrate that the GHZ states can surpass the SQL in a scaling manner, up to six photons. Since the quadratic decay behavior is quite general for short time expansion in open quantum systems (also known as the quantum Zeno effect), our results suggest a universal and scalable method to beat the SQL in the real-word metrology tasks.
\end{abstract}

\maketitle


The harness of quantum technologies revolutionizes the information processing. Quantum techonological applications has now been developed into a unprecedented stage and may greatly surpass many current approaches, typically in the domain of quantum computation \cite{divincenzo1995quantum},
quantum communication \cite{nielsen2002quantum,bouwmeester2000physics}
and quantum metrology \cite{giovannetti2004quantum}. To estimate
the unknown parameter of the embedded field, the classical strategy
of employing $N$ uncorrelated particles will yield the estimation
resolution at the particle number scaling of $\triangle^{2}\omega\propto N^{-1}$,
which, restricted by central limit theorem,  is known as the standard quantum
limit (SQL) or shot-noise limit scaling \cite{caves1981quantum,yurke19862}.
Quantum metrology indicates a more favorable solution, which leverages
the entangled states as probes, surprisingly resulting in an ultimate limit of estimation
resolution known as Heisenberg limit (HL) $\triangle^{2}\omega\propto N^{-2}$
\cite{giovannetti2004quantum,giovannetti2006quantum}. Several experimental
works have demonstrated the outperformance of quantum metrological
stategy over classical one, including diverse quantum system such
as four-photon \cite{nagata2007beating}, three trapped-ions \cite{leibfried2004toward}
and 10-spin symmetric modules \cite{jones2009magnetic}.

Entanglement is generally recognized as an crucial ingredient in
quantum metrology, and the maximally-entangled NOON state (such as Greenberger-Horne-Zeilinger (GHZ) state) is typically required for beating
the SQL. However, a quantum system will inevitably interact with surroundings.
The open system dynamics make the entangled resource gradually lose
its quantum property, eventually degrading the quantum enhancement. It is particularly demanding to investigate how to outperform the
SQL in noisy quantum metrology. It has been proven that most types
of local Markovian noise, which is usually encountered in experiment, including
losses, depolarization, spontaneous emission, both GHZ and product state could only yield an asymptotic SQL-like scaling \cite{demkowicz2012elusive,escher2011general,kolodynski2013efficient,kolodynski2014precision}. Even for the optimized choice of preparation measurement, the quantum advantages provide only constant enhancement. Nevertheless, a variety of seminal results regarding the transversal noise - that is, the direction of noise operator is perpendicular
to the Hamitonian dynamics - are presented theoretically \cite{chaves2013noisy,brask2015improved}
and experimentally \cite{zhang2019demonstrating}. Beating the SQL scaling in parameter estimation can be addressed with passive measurement by optimizing the interaction time. Moreover, provided with the Hamiltonian-not-in-Lindblad-span
(HNLS) condition hold, the Heisenberg limit scaling is attainable by actively repetitive quantum error correction (QEC) \cite{kessler2014quantum,arrad2014increasing,dur2014improved,zhou2018achieving}.
Immediately following the theoretical work, quantum metrology enhanced
by two-step QEC is experimentally demonstrated in nitrogen-vacancy
(NV) center spin register \cite{unden2016quantum}, displaying the
resilience against transversal noise.

When considering the parallel noise, e.g. the Hamiltonian dynamics
and noise operator are both along $z$-direction (dephasing noise), entanglement advantage is turned out to be an elusive goal, which hampers many developments in the field. Huelga
et al. conclude that the GHZ state and product state could both exhibit SQL-like scaling when subject to Markovian dephasing \cite{huelga1997improvement}.
 However, it is still possible to show quantum enhancement and outperform SQL in parallel noise. Later it was shown
that GHZ state can surpass the SQL, and even achieving the Zeno limit scaling $\triangle^{2}\omega \propto N^{-3/2}$ if the dephasing noise is non-Markovian with quadratic decay behaviour which
features initial quantum Zeno effect \cite{chin2012quantum,matsuzaki2011magnetic}. The entangled probes provide metrological advantages over uncorrelated probes, which is in marked contrast with Markovian noise case where a small mount of noise would restore the scaling into SQL.

In this letter, we adopt a succinct photonic method to implement the
local non-Markovian dephasing models in multipartite photonics system.
The quadratic decay factor $\gamma\left(t\right)\propto t^{2}$, which initially features the quantum Zeno dynamics, is well simulated in tabletop photonic system. By choosing the optimal interrogation time and proper measurement operator, we identify the uncertainty of parameter estimation with the probe number up to six photons. Our experimental results demonstrate that the GHZ strategy can surpass the SQL even in the presents of non-Markovian parallel noise. 

We start with recalling the quantum metrology task in the presence of independent parallel noise. With this task we aim to estimate the unknown frequency $\omega$ embedded in the unitary dynamics with $N-$qubit Hamiltonian $H=\sum_{i}^{N}H_i=\frac{1}{2}\sum_{i}^{N}\omega\sigma_{z}^{i}$,
where the $H_i=\omega\sigma_{z}^{i}$ denotes the Pauli-z operator for $i$-th
particle. The time evolution of the system (probes) operator $\rho$
in the presence of independent parallel noise is formulated with master equation
\cite{chaves2013noisy,brask2015improved}:

\begin{eqnarray}
\begin{aligned}
\frac{\partial\rho}{\partial t} & =  \mathcal{H\left(\rho\right)}+\mathcal{L\left(\rho\right)}\\ 
 & =  -i\left[H,\rho\right]+\frac{\alpha\left(t\right)}{2}\sum_{i}\left(\sigma_{z}^{i}\rho\text{\ensuremath{\sigma_{z}^{i}}}-\rho\right) \label{eq:masterequation}
 \end{aligned}
\end{eqnarray}
Here $\mathcal{H\left(\rho\right)}=-i\left[H,\rho\right]$ denotes
the unitary dynamics imprinting on the probes, and Liouvillian operator
$\mathcal{L\left(\rho\right)=}\frac{\alpha\left(t\right)}{2}\sum_{i}\left(\sigma_{z}^{i}\rho\text{\ensuremath{\sigma_{z}^{i}}}-\rho\right)$
describes the noise, of which the strength $\alpha\left(t\right)$ closely relates to the quantum non-markovianity \cite{de2017dynamics,breuer2016colloquium}. Notice that the Hamiltonian $H$ and the noise are both along the $\sigma_{z}$
direction, denoting the parallel noise. 
As the noise operator is local and independent, the effect of noise can also be well described in the single-particle system. We consider the single-particle system for simplicity. The evolution of single-qubit is formulated as $\rho(t) =\begin{pmatrix}
		\rho_{00} & \rho_{01}e^{iH_i t-\gamma(t)}\\
		\rho_{10}e^{-iH_i t-\gamma(t)} & \rho_{11}\\
	\end{pmatrix}$
where $\rho_{mn}$ are the density matrix elements of initial state. The decaying terms $e^{-\gamma(t)}$ fully characterizes the features of dephasing dynamics and is uniquely determined by interaction strength $\alpha(t)$. 
Concretely, a time-independent interaction strength $\alpha(t)$=const in eq.\ref{eq:masterequation} represents the dynamical semigroup map which is treated as typical Markovian map \cite{lindblad1976generators,alicki2007quantum}. The characteristic behavior of the quantum system exhibits the exponential decay $\gamma(t)\propto t$. The merit of our work is based on the time varied $\alpha(t)\propto t$, which leads to the quadratic behavior of $\gamma(t) \propto t^2$  that captures the basis of quantum Zeno effect and is deemed as non-markovian process.

For the uncorrelated probes, each single probe is prepared to the initial state $\left(\left|0\right\rangle +\left|1\right\rangle \right)/\sqrt{2}$,
while the entangled probes are prepared to the $N$-qubit GHZ state $\left(\left|0\right\rangle ^{\otimes N}+\left|1\right\rangle ^{\otimes N}\right)/\sqrt{2}$. The unknown frequency $\omega$ is estimated based on the average value of an operator, here we chose the parity operator
$P_{x}=\sigma_{x}^{\otimes N}$ which is optimal for GHZ probes in noiseless
case. 


\begin{figure}
\includegraphics[width=\columnwidth]{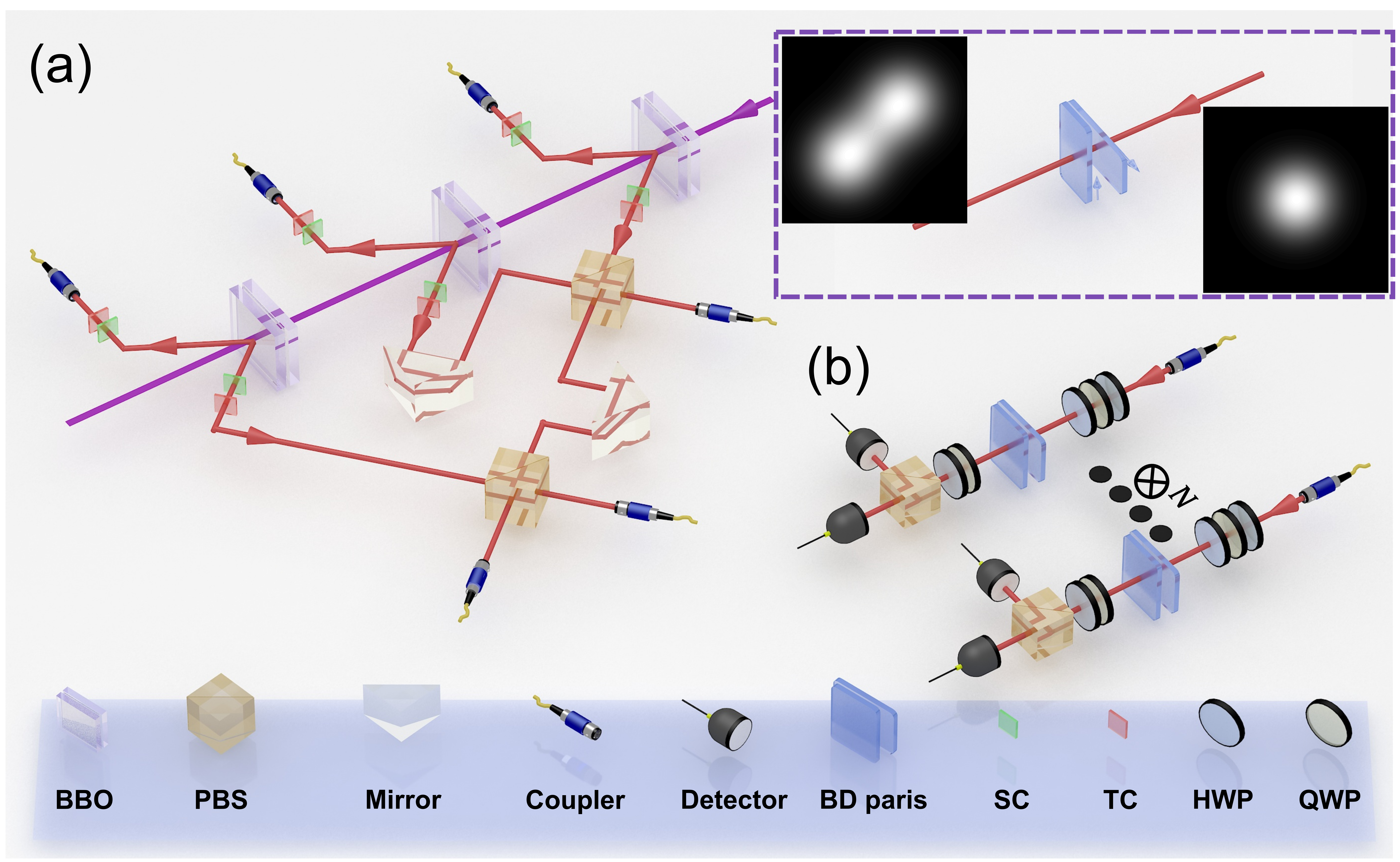}
\caption{\label{fig:setup} Sketch of the experimental setup. (a) Three ERP-pairs are generated through SPDC process with three sanwiched beam-like type II $\beta-$barium
borate (BBO) sources. The spatial compensation crystal (TC) and temporal compensation crystal (SC) make the two possible ways of generating SPDC photons with BBOs of single resource indistinguishable, so that the EPR state $\left(\left|00\right\rangle +\left|11\right\rangle \right)/\sqrt{2}$ is generated in each source. Three extraordinary photons are fed into the PBSs to complete the cascaded Hong-Ou-Mandel interferences. For N=3,4 GHZ state, the final BBO source and corresponding PBS are removed. For N=1,2, only the EPR pairs of the first BBO source is used and no HOM effect required. Then each
photon is spatially filtered by single mode fiber (SMF). (b) Halmitonian evolution and the coupling between each subsystem and environment. Finally, each photon is collected by the polarization analysis system consisting of one QWP, one HWP, one PBS and two single-photon detectors.  The HWP-QWP-HWP at $(22.5,\frac{\theta}{2},22.5)$ enables to imply the phase $e^{i\theta}$ on single party. The inset describes how the spatial mode evolved in the BD pairs. PBS: polarization beam splitter; HWP: half wave-plate; QWP: quarter wave-plate.}
\end{figure}

Fig. \ref{fig:setup} depicts the experimental setup. Three EPR pairs  are generated through Spontaneous parametric down-conversion (SPDC) by sandwiched beam-like cutting BBO sources \cite{zhang2015experimental,wang2016experimental}. The three extraordinary photons are spatially overlapped on two PBS, and become temporally indistinguishable by carefully tuning the photonic arrival time with prisms. This forms a cascaded Hong-Ou-Mandel (HOM) interferometer. Indistinguishability of the photons guarantees the occurrence of HOM effect and the GHZ states $\left(\left|0\right\rangle ^{\otimes N}+\left|1\right\rangle ^{\otimes N}\right)/\sqrt{2}$ can be generated by post-selecting the case where there is one and only one photon in each output. Afterwards, for realization of the desired dephasing channel in the following step, each photon is coupled into a single-mode fiber (SMF) and exits with the same adjustable fiber collimator to guarantee the Gaussian beam profile for the following dephasing simulator. The preparation of N-GHZ probes and spatial mode is visualized in the Fig. \ref{fig:setup}(a).

Next, each party of GHZ probes is fed into an independent
channel which comprises of an unitary dynamics $\mathcal{H\left(\rho\right)}$
and an interaction induced dephasing operator $\mathcal{L\left(\rho\right)}$ (Fig. \ref{fig:setup})(b).
Since the $\mathcal{H\left(\rho\right)}$ and $\mathcal{L\left(\rho\right)}$ are commutative, they can be imprinted independently. This could also be deduced with the expression of $\rho(t)$. As shown in Fig. \ref{fig:setup}(a), each photon first passes through a combination of  HWP-QWP-HWP which applies a dynamical phase into the GHZ states to simulate the unitary $e^{iH t}$. Then the photon undergoes a pair of thin beam-displacers (BDs) (inset in Fig.\ref{fig:setup}) to simulate the non-Markovian dephasing noise.
We encode the photonic polarization as system and the spatial mode as environment. The BD pairs consist of two BDs with identical thickness of $\ell$ and perpendicular optical axis to each other. When the photons pass through the BD pairs, the system and environment is coupled, leading to deflection along horizontal (vertical) direction according to the horizontal (vertical) polarization. 
For displacement distance $x_0/\sqrt{2}$ of each BD, the distance between different polarizations is $x_0$. The coupling leads to partial distinguishability of polarization and decoherence of the system by tracing out the environment (spatial mode). Suppose a Gaussian intensity distribution of spatial mode $\Lambda(x,y;\omega)=\frac{1}{2\pi\omega^2}e^{-\frac{x^2+y^2}{\omega^2}}$, where x(y) is Cartesian coordinates and $\omega$ is beam waist, evolution of aforementioned BD pairs suffices to imprint the quadratic decay behaviors $e^{-\gamma(t)}=e^{-\frac{x_0^2}{2\omega^2}}$. In the experiment we set $\gamma\left(t\right)=t^{2}$, and then the time evolution can  be related with the thickness $\ell$ of the BD pairs with the relation $t=\frac{x0}{\omega}, x_0=\frac{\sqrt{2}}{9.4103}\ell$. See section I of supplemental materials for detailed descriptions.

The frequency $\omega$ is estimated based on the average value of $P_x$, the mean-square error of $\omega$ can be deduced from error propagation $\triangle^{2}\omega=\left(\triangle^{2}P_{x}\right)/\left|\partial\left\langle P_{x}\right\rangle /\partial\omega\right|^{2}=\left(1-\left\langle P_{x}\right\rangle ^{2}\right)/\left|\partial\left\langle P_{x}\right\rangle /\partial\omega\right|^{2}$ (since $\langle P_x^2 \rangle=1$). If the measurement is repeated  $\upsilon$ times, the frequency resolution will also reduce by a factor of  $\upsilon$. We are interested in the performance of frequency estimation over a total time T with a single-round interrogation time t, thus $\upsilon$=T/t. The time normalized frequency resolution can be written as 
\begin{equation}
\triangle^{2}\omega T=t\frac{1-\langle P_{x}\rangle_t}{\left|\partial\left\langle P_{x}\right\rangle_t /\partial\omega\right|^{2}}
\end{equation}
which means we fix the consumption of time resource and consider only the scaling manner with respect to the particle number. In the noiseless case, the longer the evolution time, the more information can be extracted about the parameter. However, for noisy cases, there is always an optimal interrogation time to minimize Eq.5. For N-qubit GHZ probes, we can calculate the optimal interrogation time to be $t^{opt}=\sqrt{\frac{1}{4N}}$ of our noise model (see section III of Supplemental Material for details).

\begin{figure}
\includegraphics[width=1\columnwidth]{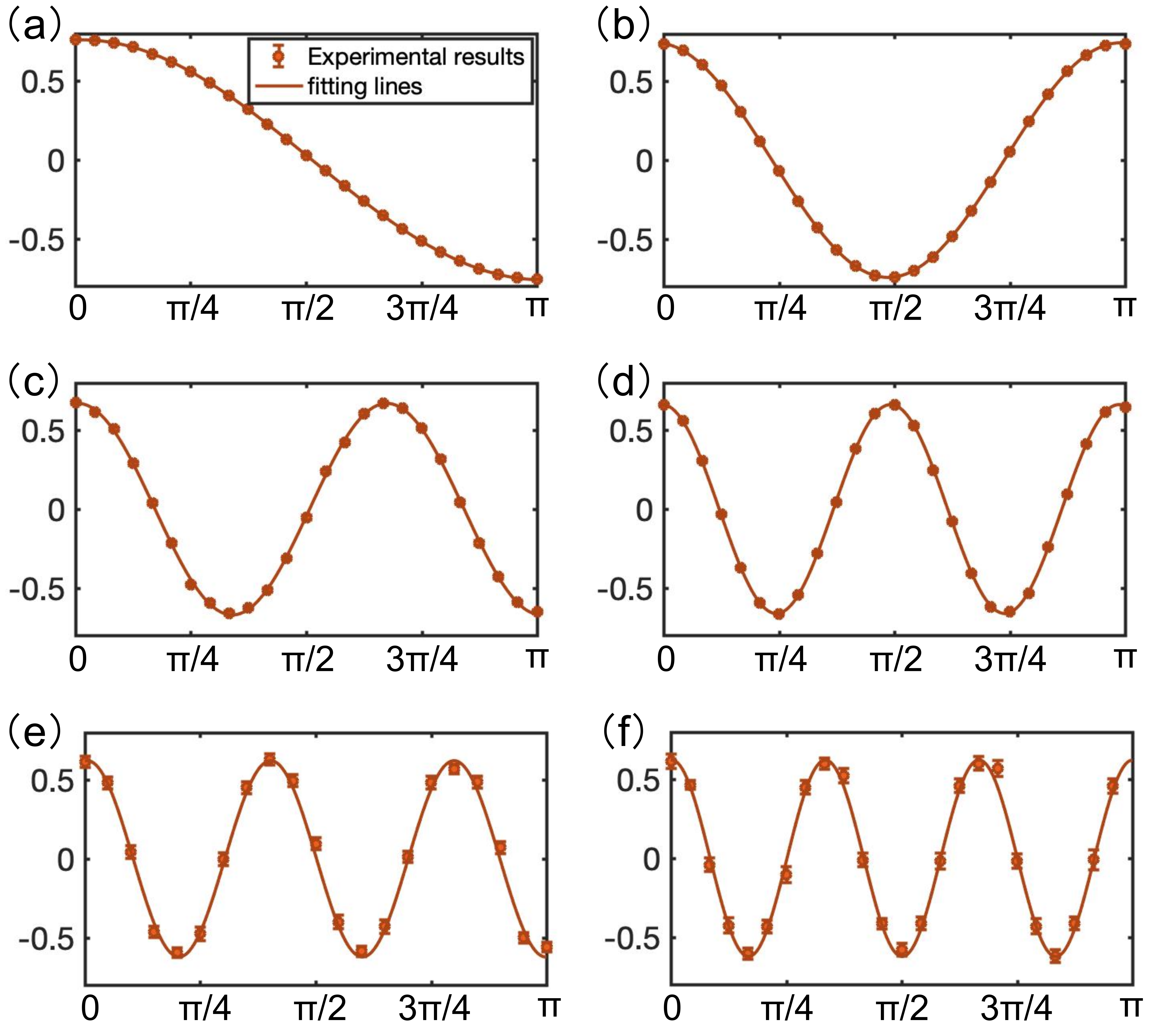}\caption{\label{fig:fringe} (a)-(f) The mean value of $P_x$ with respect to GHZ state with $N=1 \cdots 6$. The horizontal axis denotes the phase $\omega t_e$ imprinted on each subsystem. The dots represent our experimental data while the lines are the sinusoidal fitting lines. The $\frac{\partial\left\langle P_{x}\right\rangle }{\partial\omega}$ is estimated throught the derivative of fitting line at optimal interrogation point.}
\end{figure}


To experimentally measure the partial derivative, we vary the relative phase $\theta$ ($\theta=\omega t^{opt}$) for each photon from 0 to $\pi$ and get the fringe of the mean value of $P_x$. Theoretically, the $\langle P_x\rangle$ can be described with cosine function of $\langle P_{x}\rangle=\cos\left(N\omega t\right)e^{-N\gamma\left(t\right)}$. The measured data are depicted in  Fig. \ref{fig:fringe}. Each subplot denotes the fringe of N-partite GHZ state with the photon number from N=1 to 6.  We fit the data points to a cosine curve according to the theoretical evolution of $\langle P_x\rangle_t$. Next the partial derivative $\frac{\partial \langle P_x\rangle_t}{\partial \omega}$ at optimal sensitivity point ($\theta=\frac{k\pi}{2N}, k\ odd$) can be estimated. We also determine the partial derivative by using the five-point stencil method without the assumption of the evolution function of $\langle P_x\rangle_t$. The results fulfill with the curve fitting results very well (see section II of Supplemental Material for details).



\begin{figure}
\includegraphics[width=\columnwidth]{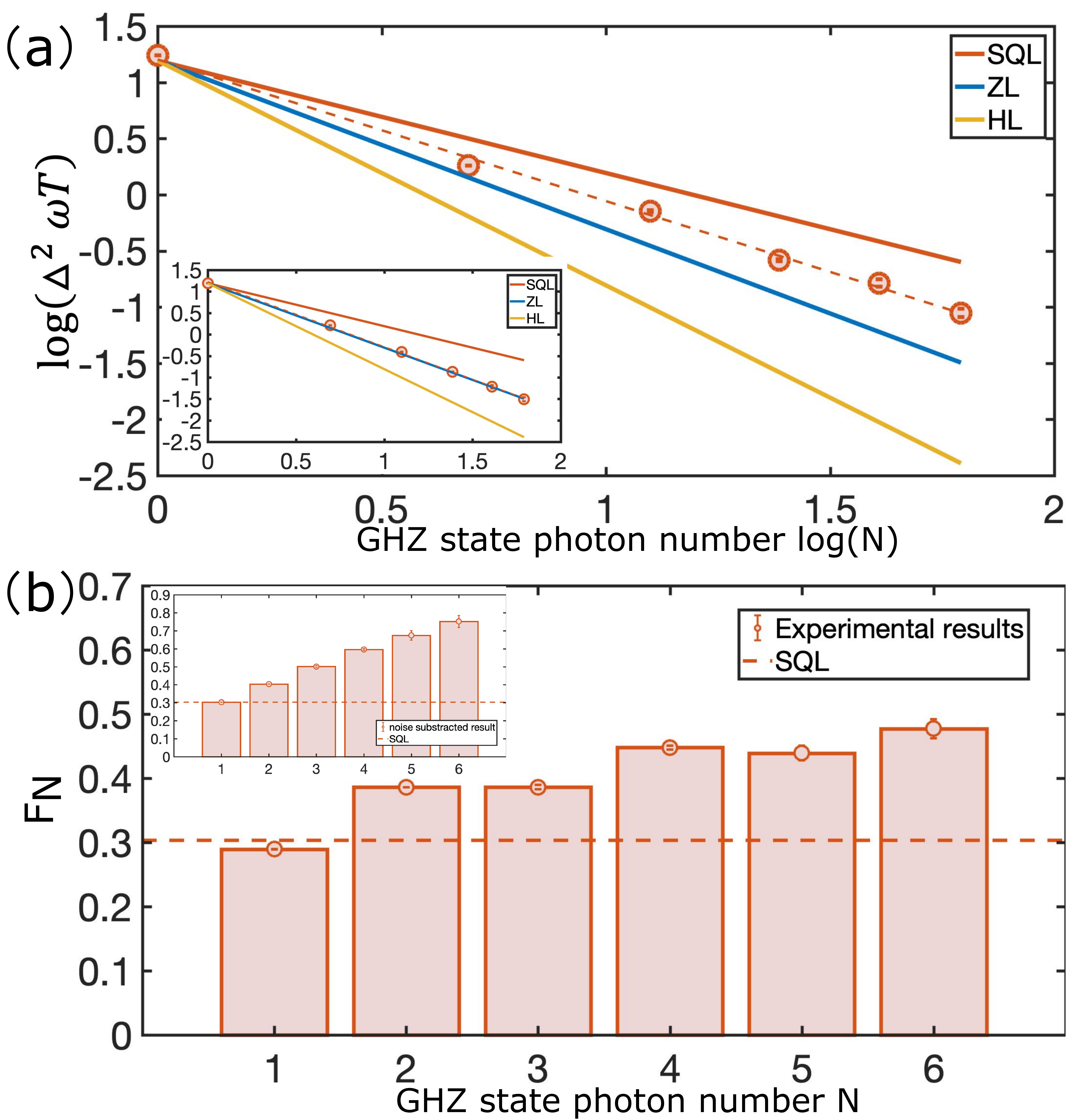}\caption{\label{fig:fisher} (a) describes the frequency resolution $\triangle^{2}\omega T$ in log(N)-log$(\triangle^{2}\omega T)$ plot, with the red, blue, yellow solid lines respectively representing the SQL, ZL and HL. The dotted line denotes the fitted line with our experimental data.  (b) The pillars shows the Fisher information per photon $1/N\triangle^{2}\omega T$
with qubit number from $N=1$ to $N=6$.  Theoretical bound of Fisher information for N=1 is presented as dotted line, in accordance with SQL. The Fisher information for $N=1 \cdots 6$ is $(0.2884\pm 0.0003,0.3861\pm 0.0001,0.3861\pm 0.0031,0.4480\pm 0.0025,0.4393\pm 0.0148,0.4774\pm 0.0155)$.  The Fisher information of all the entangled probe surpass SQL bound 0.3033. In (a)(b), the errorbars of data accounts for the Poisson-distributed of finite photon statistics, which is calculated by Monte Carlo simulations. Corresponding data with noise subtraction are provided in the inset.}
\end{figure}

Afterwards, the optimal frequency resolution $\triangle^{2}\omega T$ is calculated and displayed in the Fig. \ref{fig:fisher}(a) in logarithmic coordinate.
It shows that the results for entangled GHZ probes can beat the SQL (red line, with the slop of -1). Moreover, we calculate the Fisher information per photon  $F_N=(N\triangle^{2}\omega T)^{-1}$, which quantifies the extractable informativeness of the parameter per qubit.  The results are shown in Fig.3(b), from  which we see the quantity increases with the entangled GHZ probe size, thus indicating a scalable metrological advantage can be achieved. 
Note that due to the imperfect preparation of GHZ state, the observed frequency resolution is less than the Zeno limit $N^{-3/2}$ (ZL, the blue line) predicted by ideal entangled probes. Also, the Fisher information per photon for $N=3$ ($N=5$) cannot exceed $N=2$ ($N=3$). Because for $N=3$ ($N=5$), new EPR pair is introduced and an additional HOM interference is incorporated, which is the main noise in our preparation setup. In our experiment, we also measure the mean value of $P_x$ before evolution, the results are $\left\langle P_{x}\right\rangle _{t=0}=0.9776\pm0.0001,0.9781\pm0.0001,0.8777\pm0.0057,0.8671\pm0.0067,0.8071\pm0.0198,0.7968\pm0.0189$, for N=1 to 6 GHZ states. The $\langle P_x\rangle$ undergoes a drop when new EPR pairs are introduced. The deviation from the ideal case $\left\langle P_{x}\right\rangle_{t=0} =1$ will deteriorate the interference visibility of the fringe (or equivalently $\frac{\partial\left\langle P_{x}\right\rangle }{\partial\omega}$) and eventually degrade the resolution. We exclude the preparation imperfection by dividing the data points in Fig.2 by a factor of $\left\langle P_{x}\right\rangle _{t=0}$.  After subtracting the preparation noise, the results are shown in the insert of Fig.3. We fit the obtained resolution to a line, which yields the slop of  $-1.5126\pm0.0147$, which agrees with the Zeno limit very well. 
The increase of the GHZ probe size will inevitably lead to degrading of the performance, and finally undermines the quantum advantages. We expect that this difficulty can be overcome by future quantum error correction techniques. In section V of Supplemental Material, we analyze the bound of beating SQL with respect to the imperfect scaled probes.

The main reason the entangled probes can achieve metrological advantage in the above noise model is due to the quadratic decay behaviour $\gamma\left(t\right)\propto t^{2}$ of the noise. Such decay factor makes the entangled probes achieve their optimal interrogation time much shorter than the uncorrelated ones, and thus can experience a suppressed level of decoherence, which means they can benefit more from the non-Markovian environment. However, as mentioned in Ref.\cite{chin2012quantum,crespi2019experimental}, the short time quadratic behavior of $\gamma\left(t\right)$ is not a specific feature of our chosen noise model, but can be derived directly from a short time expansion of the Schrodinger evolution, with the only hypotheses of normalizability of the wave function and finite energy of the initial state. Such a universal decay factor is the fundamental basis of the quantum Zeno effect \cite{itano1990quantum,facchi2001quantum,fischer2001observation}. Consequently, our results indicate for almost all noise sources in real-world open quantum systems, which is the consequence of unitary evolution of the system-environment state, the SQL can be surpassed by using entangled probes with a relatively simple preparation and measuring protocol.


In conclusion, our experiment extends the scope of quantum metrology in noisy circumstance into the parallel noise scenario and demonstrates that the SQL can be surpassed under non-Markovian dephasing noise by using GHZ probes up to six qubit. The key point is to probe the system on time scales faster than the coherence time of the environment. By subtracting the preparation noise in the GHZ probes, we observe the frequency resolution fulfill the Zeno limit $N^{-3/2}$ very well, which means the precision can be improved by a factor of $N^{-1/2}$ by using entangled probes. Since the noise model investigated here is quite general, our results suggest a universal and scalable method to beat the SQL in real-world settings.
Compared to the general active noise control schemes such as quantum error correction,  the protocol we demonstrated here greatly reduces the required technology by utilizing the non-Markovian effect of the environment. Besides, the method we developed to simulate the dephasing channel could in principle simulate a broader kinds of dephasing dynamics by modulating the spatial mode of the beam. Engineering non-Markovianity has been shown beneficial for quantum communication and computation tasks \cite{bylicka2013non,laine2014nonlocal,xiang2014entanglement,liu2016efficient,huelga2012non,schmidt2011optimal}. It is an emerging perspective that to explore quantum metrology for non-Markovian processes. \cite{mirkin2020quantum,altherr2021quantum}.
It is an interesting question to investigate how metrological advantage can be achieved in a broader kinds of non-markovian environment, e.g. the dynamics with information backflow from environment to system \cite{breuer2016colloquium}, or some correlated environment between subsystems where the non-Markovianity of dynamics is not implicit locally \cite{laine2012nonlocal}.

\begin{acknowledgments}
This work was supported by the Innovation Program for Quantum Science and Technology (No. ZD0202030401), National Natural Science Foundation of China (Nos. 11734015, 11874345, 11821404, 11904357);
the Austrian Science Fund (FWF): F 7113-N38 (BeyondC), FG 5-N (Research Group);
Research Platform for Testing the Quantum and Gravity Interface (TURIS), the European Commission (ErBeSta (No.800942)), Christian Doppler Forschungsgesellschaft; 
{\"O}sterreichische Nationalstiftung f{\"u}r Forschung, Technologie und Entwicklung; 
Bundesministerium f{\"u}r Digitalisierung und Wirtschaftsstandort.
\end{acknowledgments}

\bibliography{Zenolimit}

\clearpage
\onecolumngrid 
\appendix

\section*{APPENDIX}
\section{Experimental details}

\textit{SPDC source-.} The laser with repetition rate 80 MHz and the central wavelength of 780 nm first passes through the frequency doubler to generate the 390nm pumping light for the following SPDC source. The SPDC source in our experiment is based on BBO-HWP-BBO sandwiched-like structure. Two identical 1-mm thick BBO crystals have the cutting angle for beam-like type II phase matching, which would benefit the coupling efficiency for down-converted photon. The down-converted photon is possible generated in the two BBOs, with the state $|H\rangle_{e,1}|V\rangle_{o,1}$, where the e,i(o,i) represents the extraordinary (ordinary) photons of $i$-th qubit. The polarization of photons generated in the first BBO1 will be rotated by the $45^\circ$ true zero order HWP into the state $|V\rangle_{e,1}|H\rangle_{o,1}$. Finally this two possible ways (BBO1 and BBO2) are make indistinguishable in spatial and temporal degree of freedom by the spatial and temporal compensation crystal. For a fairly comparison in our experiment from N=1 to N=6, we fix the 390 nm pumping power as 120 mw during the whole experiment, and a 3 nm (8 nm) spectral filter is applied to each e(o) photon. The corresponding counting rates for $2,4,6$-fold coincidence is about 120000 Hz, 90 Hz, and 0.07 Hz respectively.

\textit{Characterize the prepared GHZ states-.}
For the six-photon GHZ state, we also measure the measurement setting of $\sigma_z^{\otimes 6}$, accompanied with the results of $\sigma_x^{\otimes 6}$ we can deduce a lower bound on the state fidelity according to the two setting entanglement witness
\begin{equation}
    \mathcal{W}_{\mathrm{GHZ}_{\mathrm{N}}}:=3 \mathbb{I}-2\left[\frac{S_1^{\left(\mathrm{GHZ}_{N}\right)}+\mathbb{I}}{2}+\prod_{k=2}^{N} \frac{S_k^{\left(\mathrm{GHZ}_{N}\right)}+\mathbb{I}}{2}\right]
\end{equation}
where $S_1^{\left(\mathrm{GHZ}_{\mathrm{N}}\right)}:=\prod_{k=1}^{N} X^{(k)}$, $S_k^{\left(\mathrm{GHZ}_{\mathrm{N}}\right)}:=Z^{(k-1)} Z^{(k)} \quad \text { for } k=2, 3,\ldots,N.$
are stabilizers for GHZ state. The measured result is $\langle \mathcal{W}_{\mathrm{GHZ}_{\mathrm{6}}} \rangle=-0.7052\pm0.0198$, which clearly demonstrates the genuine six-photon entanglement. Then we can deduce the state fidelity $F=\langle \mathrm{GHZ}_6|\rho_{\mathrm{exp}}|\mathrm{GHZ}_6\rangle>(1-\langle \mathcal{W}_{\mathrm{GHZ}_{\mathrm{6}}} \rangle)/2=0.8526\pm0.0099$.

\textit{Simulation the non-Markovian dephasing channels-.} For simulation of desired dephasing dynamics with photonic system, we encode the polarization of photon as the system and the spatial mode as the environment. A beam displacer (BD) with deflection distance $x_0$ could couple the system and environment and lead to decoherence of system. Suppose that the optical axis of BD lies in horizontal axis, the unitary evolution of total system can be described as
\begin{equation}
   U(x_0)=|H\rangle \langle H|\otimes |x+x_0,y\rangle \langle x+x_0,y|+
   |V\rangle \langle V|\otimes |x,y\rangle \langle x,y|
\end{equation}
where $|H\rangle$($|V\rangle$) is the polarization of system and $|x,y\rangle$ spatial mode with $(x,y)$ coordinate. Another BD with axis in vertical direction correspondingly deflect the vertical polarization with the distance $x_0$ in vertical direction. 

Gnerally, provided with the initial system state $\rho(0)$ and environment state $|\phi_e\rangle$
\begin{equation}
    \rho(0) =\begin{pmatrix}
		\rho_{00} & \rho_{01}\\
		\rho_{10} & \rho_{11}\\
\end{pmatrix}, |\phi_e\rangle=\int\mathrm{d}x \mathrm{d}y \lambda(x,y;\omega)|x,y\rangle
\end{equation}
where $\omega$ is the beam waist of photon, the spatial mode intensity distribution $\Lambda(x,y;\omega)=|\lambda(x,y;\omega)|^2=\frac{1}{2\pi\omega^2}e^{-\frac{x^2+y^2}{\omega^2}}$ and satisfies $\int\Lambda(x,y;\omega)\mathrm{d}x \mathrm{d}y=1$. Coupling between the polarization and spatial mode occur when passing through the BD crystal. Considering the system alone will result in the dynamical map represented as
\begin{eqnarray}
    \rho(x_0)&=&\mathrm{Tr}(U(x_0)\rho U^\dagger(x_0)) \\ \nonumber
    &=&\begin{pmatrix}
\rho_{00} &\rho_{01}\mathcal{F}(x_0)\\
\rho_{10}\mathcal{F}^\ast(x_0) & \rho_{11}\\
\end{pmatrix}
\end{eqnarray}.
Where $\mathcal{F}(x_0)=\int \mathrm{d}x \mathrm{d}y \lambda^\ast(x,y;\omega)\lambda(x-x_0,y;\omega)=e^{-\frac{x_0^2}{2\omega ^2}}$ it Fourier transformation that describes the characteristic behavior of  dephasing dynamics. Obviously, different type of characteristic behavior could be simulated by engineering proper spatial mode amplitude distribution $\lambda(x,y;\omega)$.
    
Notice that using single BD could introduce the decoherence of polarization due to coupling with spatial mode, but this will also introduce additional decoherence by temporal distinguishability between polarizations due to deflection and short coherence time of SPDC photons. To eliminate this temporal distinguishability, we adopt a pair of thin BD pairs with the same displacement distance $x_0=y_0$ that deflect both horizontal(vertical) polarization along x(y) direction. The BD is made by the calcite crystal which the cutting angle is set as $\phi=45^\circ, \theta=90^\circ$. Relations between the thickness of BD $\ell$ and displacement distance $x_0$ of BD pairs (where each BD has a displacement distance $\frac{x_0}{\sqrt{2}}$) is calculated by $x_0=\frac{\sqrt{2}}{9.4103}\ell$.

\begin{figure}
\includegraphics[width=1\columnwidth]{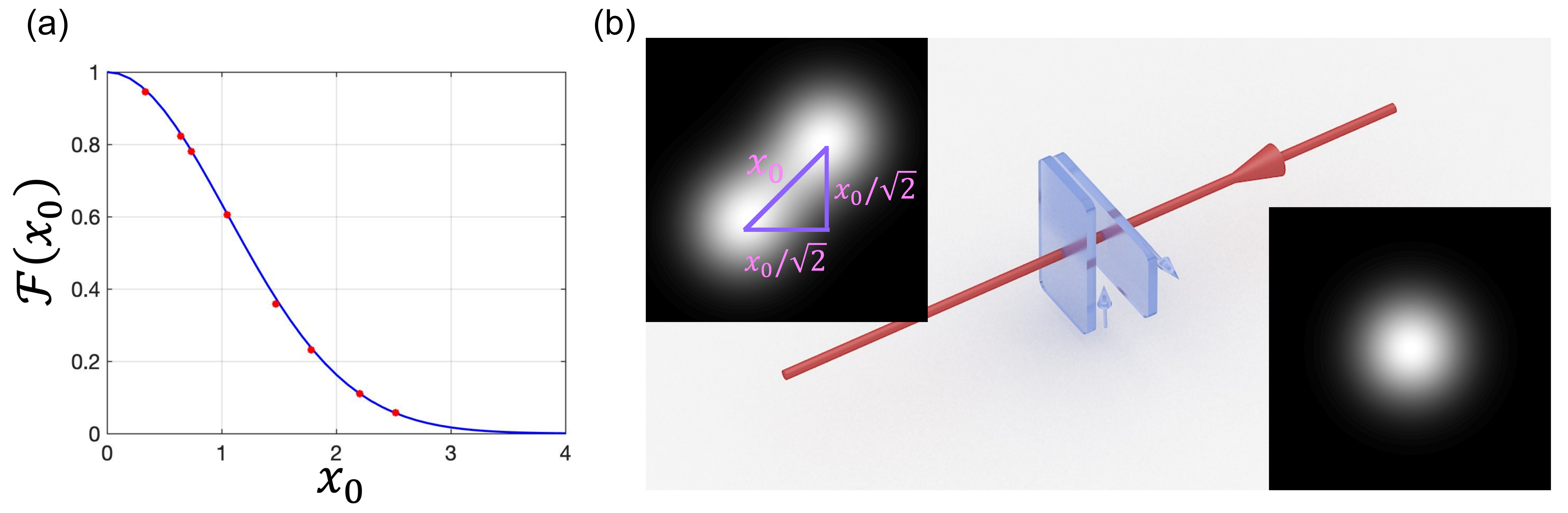}\caption{\label{fig:BD} (a)Decays behavior measured with various thickness of BD pairs and classical light. The dots denotes the tested data and the solid line is theoretical prediction. (b)sketch of the dephasing setup. The azure arrow represents the optical axis of BD. For a single BD with displacement distance $x_0/\sqrt{2}$, a pairs of BD enable to separate two polarizations with distance $x_0$.}
\end{figure}

To confirm the validity of the method, we measure the decoherence behavior of the light with respect to various displacement distance of thin BD pairs. The beam waist determined by coupler is $\omega=1.05mm$ and the initial light is set into diagonal basis $|+\rangle=(|H\rangle+|V\rangle)/\sqrt{2}$. The measured data and decay fringe is given by Table.\ref{tab:table1} and Fig. \ref{fig:BD}(a). The tested results perfectly match the theoretical prediction, which justify the validity of the simulation method.

\begin{table}
\begin{ruledtabular}
\begin{tabular}{l|cccccccc}
\textrm{Displacement distance $x_0$ of each BD}& 0.235& 0.455& 0.52& 0.74& 1.04& 1.26& 1.56& 1.78\\
\colrule
Intensity of projector $|+\rangle \langle+|$ & 4.09 & 3.91 & 3.82& 3.4& 2.9& 2.63& 2.33& 2.21\\
Intensity of projector $|+\rangle \langle+|$ & 0.116 & 0.381 & 0.471& 0.836& 1.37& 1.64& 1.87& 1.97\\
\end{tabular}
\end{ruledtabular}
\caption{\label{tab:table1}%
Relations between the thickness of BD pairs and dephasing behavior.}
\end{table}

\section{Determine the partial derivative and Fisher information via five-point stencil}

The first derivative of a function $f$ at a point $x$ can be approximated by using five-point stencil method
\begin{equation}
f'(x)\approx \frac{-f(x+2h)+8f(x+h)-8f(x-h)+f(x-2h)}{12h}.
\end{equation}
where $h$ is the grid length. The error of this approximation is of the order $h^4$, which can be obtained from Taylor expansion of the right-hand side.

We also use this method to determine the partial derivative $\frac{\partial\left\langle P_{x}\right\rangle_t }{\partial\theta}|_{t=t^{opt}}$ without the assumption of the evolution function of the mean value of parity operator $P_x$.
In the experiment, by increasing the phase of each photon from zero, the mean value of $P_x$  will undergo oscillations. Then we choose five $\theta$ (data from Fig.2 in main text) around the optimal sensitivity point $\theta=\frac{\pi}{2N}$ to calculate the partial derivative. The grid length is $\delta\theta=\{\frac{\pi}{24},\frac{\pi}{24},\frac{\pi}{24},\frac{\pi}{24},\frac{\pi}{20},\frac{\pi}{24}\}$ for $N=1-6$. The Fisher information per photon $F_N=\left(\frac{\partial\left\langle P_{x}\right\rangle_t }{\partial\theta}|_{t=t^{opt}}\right)^2 t^{opt}/N$ is calculated to be  $\{0.2938\pm 0.0042,0.3842\pm 0.0014,0.4217\pm 0.0210,0.4354\pm 0.0110,0.4996\pm 0.0295,0.5054\pm 0.0330\}$ for $N=1-6$, which coincides with the results by sinusoidal fitting very well and all beating the SQL.

\section{Derivation of the estimation sensitivity in  noise scenarios}
In this section, we aims to provide a general expressions on the frequency estimation sensitivity under the dephasing dynamics, including both the product state and GHZ state,and in the Markovoian ($\gamma(t)=\gamma_M t$) and non-Markovian case ($\gamma(t)=\gamma_N t^2$).

We start from the measurement operator $P_x=\sigma_x^{\otimes N}$ for $N-$qubit GHZ state (the N-qubit product state can be calculated with single qubit operator, but with N copies.). The mean value $\langle P_x \rangle$, uncertainty $\Delta^2 P_x$ and $\frac{\partial P_x}{\partial \omega}$ is respectively represented as
\begin{eqnarray}
    \langle P_x\rangle&=&e^{-N\gamma(t)}\mathrm{cos}(N\omega t)\\
    \Delta^2 P_x&=&\langle P_x^2\rangle-\langle P_x\rangle ^2=1-e^{-2N\gamma(t)}\mathrm{cos}^2(N\omega t)\\
    \frac{\partial P_x}{\partial \omega}&=&Nte^{-N\gamma(t)}\mathrm{sin}(N\omega t)
\end{eqnarray}
With a total available time resource $T$, it is possible to repeat the $\upsilon=\frac{T}{t}$ runs of experiment. Hence For the N-qubit GHZ state, the sensitivity of frequency estimation is calculated as
\begin{equation}
   \begin{aligned} \Delta^{2}\omega_{GHZ}&=\frac{\left(\Delta^{2}P_{x}\right)/\upsilon}{\left|\partial\left\langle P_{x}\right\rangle /\partial\omega\right|^{2}}\\ 
    &=\frac{1-e^{-2N\gamma(t)} \mathrm{cos}^2(N\omega t)}{N^2Tte^{-2N\gamma(t)}\mathrm{sin}^2(N\omega t)} \label{eq:sensitivityGHZ}
    \end{aligned}
\end{equation}
As for the N-qubit product state, it can be treated as repeating N runs of single-qubit experiment within time t, thus for time resource T, the total run of single qubit experiment is $\upsilon=N\frac{T}{t}$, and the sensitivity is
\begin{equation}
    \Delta^{2}\omega_{product}=\frac{1-e^{-2\gamma(t)} \mathrm{cos}^2(\omega t)}{NTte^{-2\gamma(t)}\mathrm{sin}^2(\omega t)} \label{eq:sensitivityproduct}
\end{equation}
Considering the optimal condition of interrogation time, that is 
\begin{eqnarray}
    &2Nt\frac{d\gamma\left(t\right)}{dt}=1 \label{eq:interrogation} \\ 
    &N\omega t=\frac{k\pi}{2}, k\;is\;odd \label{eq:phase}
\end{eqnarray}

(1)\textit{Markovian case.} In this case, we consider the semigroup dynamics $\gamma(t)=\gamma_M t$ raised in the original theorem paper \cite{huelga1997improvement,chin2012quantum}, where $\gamma_M$ is constant factor. 

For N-qubit GHZ state, the optimal time can be obtained to be $t_e=\frac{1}{2N\gamma_M}$. For product state, the optimal time is $t_p=\frac{1}{2\gamma_M}$. By substituting the optimal time and eq. \ref{eq:phase} into eq.\ref{eq:sensitivityGHZ} and eq.\ref{eq:sensitivityproduct} respectively, we get the ultimate expression of sensitivity of GHZ state and product state
\begin{equation}
    \Delta^{2}\omega_{GHZ}=\Delta^{2}\omega_{product}=\frac{2e\gamma_M}{NT}\\
\end{equation}

(1)\textit{non-Markovian case.} We consider the quadratic decay behavior $\gamma(t)=\gamma_Nt^2$ which has quantum Zeno effect in initial short time scale. For N-qubit GHZ state, the optimal time is $t_e=\sqrt{\frac{1}{4N\gamma_N}}$. For product state, the optimal time is $t_p=\sqrt{\frac{1}{4\gamma_N}}$. In this case, sensitivity of GHZ state and product become different
\begin{equation}
    \begin{aligned}
\Delta^{2}\omega_{GHZ}&=\frac{2(e\gamma_N)^{0.5}}{N^{\frac{3}{2}}T}\\
\Delta^{2}\omega_{product}&=\frac{2(e\gamma_N)^{0.5}}{NT}
    \end{aligned}
\end{equation}

\section{Comparison of quantum metrology in non-Markovian environment with Markovian environment}
We have demonstrated surpassing the SQL in parallel noise with quadratic-like decay of non-Markovian circumstance. Entanglement is commonly recognized as a central part to this task. But entanglement is not the only ingredient responsible to the outperformance. It is still interesting to investigate the role of non-Markovianity. As a compliment to our experiment, we compare the performance of entangled probes in non-Markovian and Markovian dephasing cases.

In the main text, we performed the non-Markovian case with $\gamma(t)=t^2$, here the pure Markovian dephasing where the exponential decay $\gamma(t)=e^{-0.5}t$. These two cases exhibit the same precision with $N=1$ but differ as the particle number increasing. For simulation of Markovian dephasing dynamics, the BD pairs is replaced by the wave-plates which is used to introduce the phase-flipped operation. The Markovian dephasing dynamics can be experimentally engineered by a proper mixture of a pure initial state
and its phase-fliped state with appropriate factions of total measurement time, then tracing out the classical time information about which state is prepared. Such a
method is generally used in optical channel simulation \cite{ringbauer2018certification}.

\begin{figure}
\includegraphics[width=0.5\columnwidth]{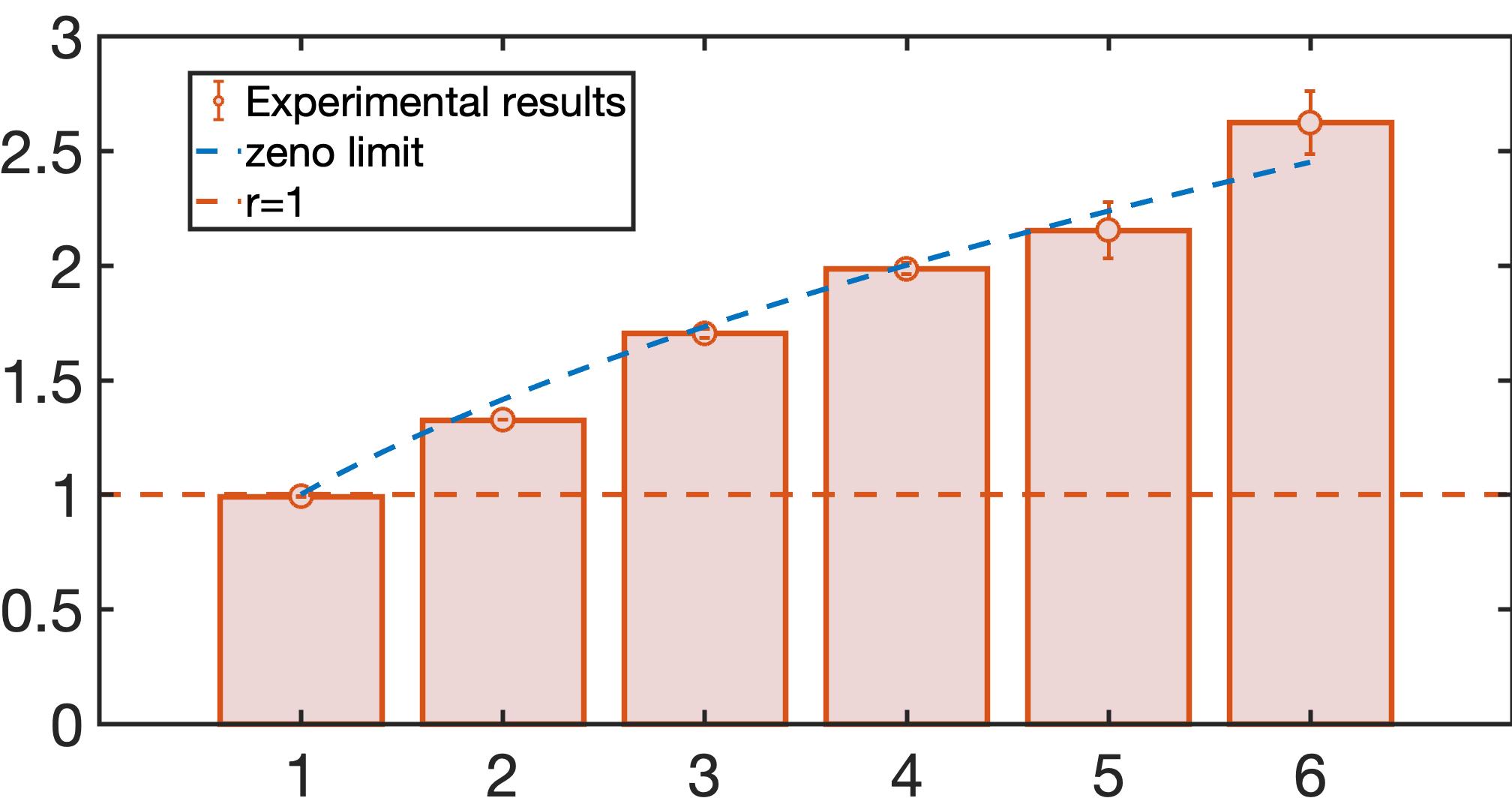}\caption{\label{fig:relative} Relative resolution of non-Markovian over Markovian case. The $r=1$ means there is no relative advantages for non-Markovian environment, which is the case when N=1.}
\end{figure}

We define the relative resolution of non-Markovian over Markovian case as
\begin{equation}
r^{2}=\frac{\left|1/\triangle^{2}\omega\right|_{NM}}{\left|1/\triangle^{2}\omega\right|_{M}}=N\left(\frac{t_{NM}}{t_{M}}\right)e^{2\gamma\left(t_{M}\right)-2N\gamma\left(t_{NM}\right)}
\end{equation}
Here the subscript $NM\left(M\right)$ denotes the non-Markoivan (Markovian)
case and the interrogation time is optimized according to eq.\ref{eq:interrogation}. Fig.
 depicts the experimental result of relative
resolution of GHZ state with $N$ ranging from 1 to 6, which accord
with the Zeno limit $r^{2}=1/\sqrt{N}$ intermediate with SQL and
HL. The reason is that in the ideal case,
that is, perfect GHZ probes preparation, the GHZ state in non-Markovian
dephasing is an alternative to the product state in Markovian dephasing
because of the same sensitivity. Thus, comparision with the these two
cases also to some degree reflects the Zeno scaling, where the preparation
imperfection of experiment could be justifiably excluded. It is worth noting that for the specific two fixed Markovian and non-Markovian environment, even if they shows the same precision in case of $n_{0}$- qubit probes, when moving toward higher qubit $N>n_{0}$, the non-Markovian one will show the advantages over Markovian one.

\section{Numerical analysis with state preparation noise}
It is a common issue that performance of quantum tasks will degrade with the increasing noise. The preparation imperfection is one of the inherent noise. Generally, preparation of entangled state is based on fusing different particles, which make preparation noise greatly increased with the entangled particle number scaled due to the increasing circuit depth. In this section, we will briefly analysis a realistic scenario where the preparation noise is scaled as additional qubit is incorporated, and calculate the performance of quantum metrology in non-Markovian dephasing dynamics.

\begin{figure}
\includegraphics[width=0.5\columnwidth]{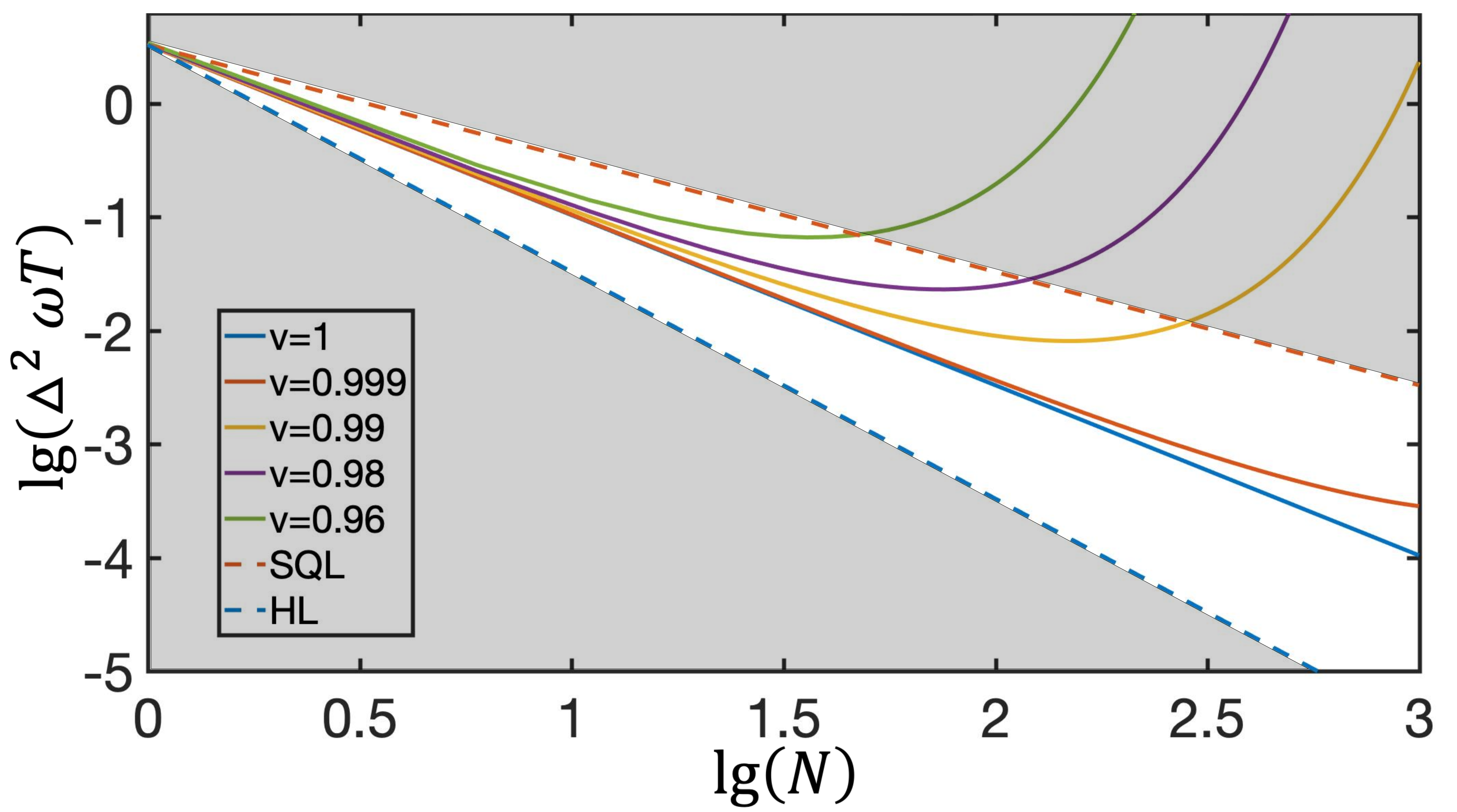}\caption{\label{fig:preparationnoise} Numerical analyasis of relations of the sensitivity with respect to qubit number N up to $10^3$. The upper and down shaded areas represent the SQL and HL respectively.}
\end{figure}

For the SPDC photons, extending the entangled number of particles by two requires one fusion gate, thus it is reasonable to model the visibility of prepared state with a exponentially decay with numbers of fusion gate. We consider the white noise for simplicity 
\begin{equation}
    \rho_{noise}=v^{N/2}|\mathrm{GHZ}\rangle\langle \mathrm{GHZ}|_N+\frac{1-v^{N/2}}{2^N}\mathcal{I}
\end{equation}
where the $|\mathrm{GHZ}>=(|0\rangle^{\otimes N}+|1\rangle^{\otimes N})/\sqrt{2}$ and $\mathcal{I}$ is white noise. The visibility for N-qubit GHZ state is $v^{N/2}$. In Fig. \ref{fig:preparationnoise}, we numerically calculate the performance of quantum metrology with respect to the particle number, under the various condition of visibility. It is immediately clear that any small imperfection of fusion leading to imperfect visibility will eventually destroy the advantages of superclassical prevision when the N is sufficient large. However, the SQL can be surpassed when in moderate mount of qubit number and visibility imperfection, which is within the reach of current quantum technology.
Overcoming the imperfection accumulation as the circuit depth increasing is a notoriously difficult in quantum technology. In the future, A through investigation of non-Markovianity and accompanying method to take use of the non-Markovian features to reduce the detrimental effect of noise is expected to applied on the quantum metrology.

\end{document}